\documentclass[a4paper,11pt]{article}
\pdfoutput=1 

\usepackage{jheppub} 

\usepackage[T1]{fontenc}

\title{\boldmath On the entropic nature of unified interactions}

\author[a,1]{Emre Dil\note{Corresponding author.}}
\author[b,c]{and Tugrul Yumak}

\affiliation[a]{Department of Energy Systems Engineering, Beykent University,\\Sariyer, 34000 Istanbul, Turkey}
\affiliation[b]{Department of Mechanical and Aerospace Eng., West Virginia University, Morgantown, 26506-6101 West Virginia, USA}
\affiliation[c]{Department of Chemistry, Sinop University,\\Korucuk, 57000 Sinop, Turkey}

\emailAdd{emredil@beykent.edu.tr}
\emailAdd{tugrul.yumak@mail.wvu.edu}

\abstract{In this paper, we generalize Verlinde's entropic gravity proposal on the other fundamental interactions of nature. We begin by introducing the entropic origin of the Coulomb's electrostatic force, and then the magnetic force, by assuming the holographic principle holds when a charged particle approaches a screen enclosing the emerged part of the spacetime due to a source. We assume that the entropy of the screen changes when the charge approaches as in the Verlinde's approach. Thereafter, we obtain the entropic Maxwell equations in both classical and covariant form by means of the holographic principle hold for the source. Considering the gauge covariant structure of the fundamental forces, we implicitly generalize the entropic origin of the electromagnetic force into the strong and weak nuclear forces. The possibility of the entropic origin of the all fundamental forces of nature is assumed to make a meaningful contribution to the unification scheme of the fundamental forces.}

\begin{document} 
\maketitle
\flushbottom

\section{\label{sec:level1}Introduction}

Gravity has begun to be considered as an entropic force rather than a fundamental force with the pioneering work of Verlinde and others ~\cite{a,b1,b,c,d}. According to the Verlinde's proposal, gravity and spacetime can emerge from a microscopic description whose specific details are independent from the general principles of the theory. The main effect creating the gravity is the amount of information measured as entropy due to the displacement of matter. Reaction force to the change of entropy during the displacement of the matter leads to the gravitational force.

The information stored in a spacetime emerged by a source matter obeys the holographic principle that a finite number of degrees of freedom (bits) is associated with the information in the volume. The number of bits on the holographic screen enclosing the emergent part of the spacetime is proportional to the surface area of the holographic screen ~\cite{e,f}. The total energy resulting from the source mass is evenly distributed over the bits on the holographic boundary on which the source leads a certain temperature. On the other hand, the main motivation of Verlinde's proposal is related to the Bekenstein's assumption ~\cite{g} on the entropy formula. Bekenstein argued that a Planck mass approaches to the horizon of a black hole by one Planck length and it leads to an increase of one bit of information, as $2\pi k$, on the surface area of the horizon. Similar to the that Verlinde proposed a mass approaching a holographic screen leads to the same amount of entropy change on the screen. Consequently, the product of the change of entropy and the temperature on the screen gives the work done by the emergent gravitational force, in Verlinde's proposal.

Verlinde obtained the Newton's law of motion and the gravitational field equations in the framework of entropic origin approach and this has motivated Freund to ask whether the electromagnetic force and the other fundamental forces can be described as a force having an entropic origin ~\cite{h}. He could qualitatively extend the entropic origin of gravity into abelian and non-abelian gauge fields. In order to bring a quantitative description for Freund arguments, here we firstly begin with electrostatic force and apply Verlinde's holographic argument on a charge based entropy change and energy creation on the boundary screen of the emergent spacetime due to another source charge. After obtaining the electrostatic force and Coulomb's force equation, we proceed to get the magnetic field equation for a charge current. We then generalize the entropic origin of electromagnetic field equations to the classical and covariant form of Maxwell equations as abelian gauge field equations in order to construct a bridge for obtaining the entropic origins of the non-abelian strong and weak field interactions in gauge field theory structure. The success of the improving entropic origination of all the fundamental forces of nature is expected to make a significant contribution to the unification studies of the four fundamental forces, in the physics community.

\section{\label{sec:level2}Entropic origin of electric and magnetic forces}
Entropic origin of gravity leads to the investigations on whether the other forces of nature can be originated as an entropic force in the literature ~\cite{h,i,j}. Although it is an important approach, only a few leading studies attempts to investigate the entropic origins of the electromagnetic force. The most significant one in ~\cite{j} tries to generalize electromagnetic interaction with the steps taken by Verlinde, and there resulted a bizarre impairing in the physical significance of the formal thermodynamical interpretation with the existence of a negative electromagnetic temperature.

In this study, we overcome this inconsistency by repairing the description of the energy on the holographic screen due to the source charge. We begin with the assumption that a test particle of Planck charge $q_p$ approaches to the holographic screen from a direction at which spacetime has already emerged due to a source charge $q_2$ which can even be written in terms of the Planck units. Similar to Bekenstein-Verlinde argument, the change of entropy on the holographic boundary of the emerged spacetime is $2\pi k$, as the test charge approaches to the screen by one Compton wavelength of Planck length $l_p$:
\begin{equation}
\label{a}
\Delta S = 2\pi k \,,
\qquad
q_p = \sqrt{4\pi \epsilon_0 \hbar c} \,,
\end{equation}
where we used the Planck charge expression for the test charge $q_p$. Accordingly, the thermodynamical work done on the test charge due to the entropy change during the displacement is
\begin{equation}
\label{b}
\begin{split}
Fl_P &= T \Delta S \,
\\
&= T 2\pi k \frac{q_p} {\sqrt{4\pi \epsilon_0 \hbar c}} \,.
\end{split}
\end{equation}
Stated that the source charge $q_2$ leads to the emerging of spacetime which is bounded by the holographic screen and a temperature on it, this temperature makes the test particle gain an acceleration with the Unruh law:
\begin{equation}
\label{c}
kT = \frac{1}{2 \pi} \frac{\hbar}{c} a \,.
\end{equation}
Substituting the Unruh temperature \eqref{c} into \eqref{b} gives
\begin{equation}
\label{d}
\begin{split}
F &= \frac{T 2\pi k}{l_p} \frac{q_p} {\sqrt{4\pi \epsilon_0 \hbar c}} \,
\\
&= \frac{\hbar a}{c l_p} \frac{q_p} {\sqrt{4\pi \epsilon_0 \hbar c}} \,
\\
&= \frac{\sqrt{\hbar c}}{{l_p}^{2}} \frac{q_p}{\sqrt{4 \pi \epsilon_0}} \,.
\end{split}
\end{equation}
where we use $a=c/t_p$ and $l_p=ct_p$ in the second line in terms of the Planck units. Note that the source charge $q_2$ leads to the temperature on the holographic screen, which gives raise to the first factor in the last line. Using Planck charge expression $q=\sqrt{4\pi \epsilon_0 \hbar c}$ for $q_2$ \eqref{d} turns out to be
\begin{equation}
\label{e}
F = \frac{q_2}{4 \pi \epsilon_0 l_p^2} q_p = q_p E \,.
\end{equation}
It is the electrostatic force on a test charge in an electric field $E$ produced by the change of information on the emergent holographic spacetime boundary, due to a source charge leading to a temperature on it.

\subsection{\label{sec:level3}Coulomb's force}
We now proceed to repair the inconsistency in ~\cite{j} while describing the total energy on the emergent spherically symmetric spacetime boundary. Since Verlinde considers the gravity, he takes the $Mc^2$ energy into account as the causing effect for the temperature $T$ on the holographic boundary. However, Wang makes a direct analogy by replacing the charge $Q$ with the mass $M$, as the energy source, and takes $Qc^2$ into account. however, this quantity does not define an energy on the holographic screen. We, therefore, repair this inconsistency by taking the Planck energy $\hbar/t_p$ as the source energy due to the charge in Planck units and this energy results a temperature $T$ on the screen as
\begin{equation}
\label{f}
E_p = \frac{\hbar}{t_p} = \frac{1}{2} NkT \,,
\end{equation}
where $N$ is the total number of bits on the spherically symmetric holographic screen, and it is again proportional to the surface area of the screen, such that
\begin{equation}
\label{g}
N = \frac{A}{c^2 {t_p}^2} \,.
\end{equation}
Since power in Planck units is $\hbar/{t_p}^2 = c^5/G$ for the source charge $q_2$, ~\eqref{f} can be rewritten as
\begin{equation}
\label{h}
q_2 \sqrt{\frac{c^4}{4 \pi \epsilon_0 G}} = \frac{1}{2} NkT \,,
\end{equation}
and
\begin{equation}
\label{i}
q_2 \sqrt{\frac{\hbar}{4 \pi \epsilon_0 c {t_p}^2}} = \frac{1}{2} kT \frac{A}{c^2 {t_p}^2} \,,
\end{equation}
which yields the temperature on the screen, then
\begin{equation}
\label{j}
T = \frac{2 q_2}{kA} \sqrt{\frac{\hbar c^3 {t_p}^2}{4 \pi \epsilon_0}} \,.
\end{equation}
Now we can obtain the force on a test particle with charge $q_p$ approaching this spherically symmetric spacetime boundary emerged due to the source charge $q_2$ by a distance of one Compton wavelength as Planck length $l_p$. Then, the entropy change on the screen is again $2 \pi k$ by Bekenstein-Verlinde argument, and the thermal work done on the test particle in terms of the entropy change is
\begin{equation}
\label{k}
Fl_p = T \Delta S = \frac{2 q_2}{kA} \sqrt{\frac{\hbar c^3 {t_p}^2}{4 \pi \epsilon_0}} 2\pi k \frac{q_p} {\sqrt{4\pi \epsilon_0 \hbar c}} \,,
\end{equation}
where we have used the screen temperature ~\eqref{j}. Simplifying ~\eqref{k} gives
\begin{equation}
\label{l}
F = \frac{1}{4 \pi \epsilon_0} \frac{q_p q_2}{{l_p}^2} \,,
\end{equation}
which is obviously the Coulomb's force originating from the entropy change on a holographic screen due to the displacement of a test charge under the influence of the temperature produced by the source charge.

\subsection{\label{sec:level4}Biot-Savart law}
For this case, we assume that the source producing the energy and temperature on the emergent spacetime boundary is a charge current, and we attempt to obtain the force on another charge current close to the boundary of the spacetime. Therefore, the temperature of the screen due to the source current is obtained by reinterpreting ~\eqref{h} with the fact that $\epsilon_0 \mu_0 = c^{-2}$, such as
\begin{equation}
\label{m}
q_2 \sqrt{\frac{\mu_0}{4 \pi} \frac{\hbar c}{{t_p}^2}} = \frac{1}{2} kT \frac{A}{c^2 {t_p}^2} \,,
\end{equation}
which gives
\begin{equation}
\label{n}
T = \frac{2 I_2}{kA} \sqrt{\frac{\mu_0}{4 \pi} \hbar c^5 {t_p}^4 } \,.
\end{equation}
where we have used $I_2 = q_2/t_p$ for the source current. Then the entropy change due to the approaching test current $I_1$ leads to the work,
\begin{equation}
\label{o}
Fl = T2 \pi k I_1 \sqrt{\frac{\mu_0}{4 \pi} \frac{c {t_p}^2}{\hbar}} \,.
\end{equation}
Using ~\eqref{m} here, we obtain
\begin{equation}
\label{p}
F = \frac{\mu_0}{4 \pi} \frac{I_2l}{r^2} I_1l = B I_1 l \,.
\end{equation}
and
\begin{equation}
\label{ppp}
B = \frac{\mu_0}{4 \pi} \frac{I_2l}{r^2} \,.
\end{equation}
Here, ~\eqref{ppp} is the Biott-Savart law and $B$ is apparently the magnitude of the magnetic field due to the current $I_2$ and inferred from the factor standing in the middle of ~\eqref{p}.

As it is clear that the Planck scale is only valid for the quantum world, but here we only assume that the quantities in classical Biott-Savart law are converted into Planck units as conventional unit conversions. This approach has also been followed by Verlinde in his entropic gravity study ~\cite{a} as offering to assume the classical quantities in gravitational law equations to be converted into Planck units as in the conventional unit conversion method. This does not mean to treat the macroscopic objects as quantum objects. This does just mean to mimic the conventional unit conversion method.

\section{\label{sec:level5}Derivation of Maxwell's equations through entropic origin}

We firstly begin with the derivation of Gauss's law, by assuming the source charge causes the thermal energy which is randomly divided over the $N$ bits on the spherical holographic screen enclosing the emergent part of the spacetime:
\begin{equation}
\label{q}
E = \frac{1}{2} \int{kT dN}
\end{equation}
Using Planck power in order to convert $E$ in terms of the electrostatic energy and Unruh law gives
\begin{equation}
\label{r}
q_2 \sqrt{\frac{\hbar}{4 \pi \epsilon_0 c {t_p}^2}} = \int{\frac{\hbar}{4 \pi c^2 {t_p}^3} dA} \,,
\end{equation}
where $dN = dA/c^2 {t_p}^2$ is used. We then write the charge in terms of the charge density, such as
\begin{equation}
\label{s}
\int{\frac{\rho}{\epsilon_0} dV}  = \int{\sqrt{\frac{\hbar c}{4 \pi \epsilon_0 c^4 {t_p}^4}} dA} \,.
\end{equation}
Applying the divergence theorem on the right hand side of ~\eqref{s} leads to
\begin{equation}
\label{t}
\begin{split}
\int{\frac{\rho}{\epsilon_0} dV}  &= \int{\nabla . \left(\sqrt{\frac{\hbar c}{4 \pi \epsilon_0 l^4}} \hat{l} \right) dV} \,
\\
 &= \int{\nabla . \left(\frac{q_2}{4 \pi \epsilon_0 l^2} \hat{l} \right) dV} \,.
\end{split}
\end{equation}
This is obviously equivalent to the Gauss's law as
\begin{equation}
\label{v}
\nabla . E = \frac{\rho}{\epsilon_0} \,.
\end{equation}

Similarly, one can obtain the emergent Ampere's law for a source of current producing a temperature on the holographic screen due to the magnetic energy which is again divided over the $N$ bits, as follows
\begin{equation}
\label{w}
E = q_2 \frac{\mu_0}{4 \pi} Ic = \frac{1}{2} \int{kT dN}
\end{equation}
where $\mu_0 Ic/4 \pi$ is the impedance due to the source current $I = q_2/t$. Rewriting the Unruh temperature on the left hand side and the current in terms of the surface current density gives
\begin{equation}
\label{x}
\int{\frac{q_2 c}{4 \pi} \mu_0 J . dA} = \int{\frac{\hbar}{4 \pi c^2 {t_p}^3} dA} \,,
\end{equation}
and rearranging ~\eqref{x} yields
\begin{equation}
\label{y}
\int{\mu_0 J . dA} = \int{\left(\frac{\mu_0}{4 \pi} \frac{Il'}{l^3} \hat{n}\right) . dA} \,,
\end{equation}
where $\hat{n}$ is parallel to $dA$. The term in the parenthesis can be assumed as the rotational of a vector, such as
\begin{equation}
\label{z}
\int{\mu_0 J . dA} = \int{\nabla \times \left(\frac{\mu_0}{4 \pi} \frac{Il'}{l^3} \hat{n}_\bot \right) . dA} \,.
\end{equation}
We obviously reach the emergent Ampere's law, since the term in parenthesis is the magnetic field due to the source current, so
\begin{equation}
\label{a1}
\nabla \times B = \mu_0 J \,.
\end{equation}
Combining the processes to obtain the results in ~\eqref{v} and ~\eqref{a1} leads to the Maxwell equations. To proceed, we express the energy on the holographic screen enclosing the emergent part of the spacetime due to the source charge $q_2$ inspired by ~\eqref{r}, as follows
\begin{equation}
\label{b1}
q_2 = \int{\frac{\hbar c}{4 \pi l^3 \phi} dA} \,,
\end{equation}
where $\phi$ is the potential due to the source charge. In order to obtain a Lorentz covariant relativistic form of the emergent field equations, we need to combine equation ~\eqref{b1} both for a rest charge and a current, such that
\begin{equation}
\label{c1}
\frac{q_2}{\epsilon_0} = \int{\frac{1}{4 \pi \epsilon_0} \frac{\hbar c}{\phi l^3} dA} \,,
\end{equation}
\begin{equation}
\label{d1}
\mu_0 I = \int{\frac{\mu_0}{4 \pi} \frac{\hbar c}{\phi l^3 t_p} dA} \,.
\end{equation}
Combining these separate equations into one single covariant form means that there occurs one time-like and one space-like components of the same equation. For a 4-current $I^{\mu} = (cq_2, I)$, we can combine them as
\begin{equation}
\label{e1}
\mu_0 I^0 = \int{F_{\alpha}^0 dA^{\alpha}} \,,
\end{equation}
\begin{equation}
\label{f1}
\mu_0 I^i = \int{\partial_{\alpha}{F^{\alpha i}} dA} \,,
\end{equation}
where $F^{\alpha \beta}$ is the electromagnetic field strength tensor which corresponds to the emergent field, as follows
\begin{equation}
\label{g1}
\mu_0 I^0 = \int{\mu_0 c \rho dV} = \int{\partial_{\alpha}{F^{\alpha 0}} dV} \,,
\end{equation}
\begin{equation}
\label{h1}
\mu_0 I^i = \int{\mu_0 J^i dA^i} = \int{\partial_{\alpha}{F^{\alpha i}} dA^i} \,.
\end{equation}
Bringing them into a single covariant form leads to the Maxwell equations, such that
\begin{equation*}
\left(\nabla . \left[\frac{1}{4 \pi \epsilon_0} \frac{\hbar c}{\phi l^3}\hat{l}\right], \left[\frac{\mu_0}{4 \pi} \frac{\hbar c}{\phi l^3 t_p}\hat{n}_\bot \right] \right) = \left( \partial_{\alpha}{F^{\alpha 0}}, \partial_{\alpha}{F^{\alpha i}} \right)
\end{equation*}
\begin{equation}
\label{i1}
\left(\frac{1}{4 \pi \epsilon_0} \frac{3\hbar c}{\phi l^4},\frac{\mu_0}{4 \pi} \frac{\hbar c}{\phi l^3 t_p}\hat{n} \right) = \partial_{\alpha}{F^{\alpha \beta}} \,,
\end{equation}
then substituting this into ~\eqref{g1} and ~\eqref{h1}, we obtain
\begin{equation}
\label{j1}
\mu_0 J^{\beta} = \partial_{\alpha}{F^{\alpha \beta}} \,,
\end{equation}
which is the inhomogeneous Maxwell equations. Since the electromagnetic interaction is a gauge field theory like the strong and weak interactions as the fundamental interactions of nature, we can infer the emergent nature of these two forces by following the similar steps taken for the electromagnetic interaction.

\section{\label{sec:level6}Entropic nature of the nuclear forces}

The main difference between electromagnetic and nuclear forces is the non-abelian nature of the gauge fields of nuclear forces, while it is abelian for the electromagnetic force. Despite the gauge symmetry of strong and weak nuclear forces differs as $SU(3)$ and $SU(2)$, respectively, this does not change the fundamental entropic structure of the interactions. The similarity of the gauge structure of these three fields yields proposing the existence of entropic origin for two nuclear forces, once the existence of entropic origin is clearly expressed for the electromagnetic one.

We need to define the strong and weak charges in Planck units to proceed. It is known that the electric charge is defined in terms of the coupling strength (fine structure constant) $\alpha_e$ in Planck units as $q_p = \sqrt{4 \pi \epsilon_0 \hbar c} = e / \sqrt{\alpha_e}$. From this fact the strong and weak charges are
\begin{equation}
\label{k1}
q_s = \frac{g_s}{\sqrt{\alpha_s}} \,,
\qquad
q_w = \frac{g_w}{\sqrt{\alpha_w}} \,,
\end{equation}
Here, the coupling strength values for the strong and weak interactions are $\alpha_s = g_{s}^{2} / 4\pi$ and $\alpha_w = g_{w}^{2} / 4\pi$, respectively, which makes ~\eqref{k1} turn out to be
\begin{equation}
\label{l1}
q_s = q_w = \sqrt{4 \pi} \,,
\end{equation}
We then assume that a source charge $q_{sw}$ leads to a thermal energy distributed over the $N$ bits on the holographic screen enclosing the emergent part of the spacetime due to the existence of the source. Accordingly, the thermal energy can be expressed in terms of a nuclear potential $\mathcal{V}_{sw}$, as
\begin{equation}
\label{m1}
\begin{split}
E = q_{sw} \mathcal{V}_{sw} &= \frac{1}{2} \int{kT dN} \,
\\
&= \int{\frac{1}{4 \pi} \frac{\hbar c}{l^3} dA} \,,
\end{split}
\end{equation}
and
\begin{equation}
\label{n1}
q_{sw} = \int{\frac{1}{4 \pi} \frac{\hbar c}{\mathcal{V}_{sw} l^3} dA} \,.
\end{equation}
Since there is no permittivity or permeability terms for strong and weak interactions, from the symmetry of the structures of the gauge field theories we postulate that ~\eqref{n1} must directly lead to the field equations of nuclear forces, such that
\begin{equation}
\label{o1}
D_{\mu} G^{\mu \nu} = g_{sw} J^{\nu} \,,
\end{equation}
where $G^{\mu \nu}$ is the field-strength tensor for the corresponding nuclear force. The strong and weak interactions are based on a non-abelian gauge field theory, which couple minimally to the matter fields. Denoting the fermionic matter field with $\psi$, the interaction term reads in the Lagrangian as $g_{sw}\overline{\psi}\gamma^{\mu}q_{sw}\psi \mathcal{V}_{\mu}$. Here, the term in the interaction Lagrangian includes the 4-current density $J^{\mu}=\overline{\psi}\gamma^{\mu}q_{sw}\psi$  associated with the gauge field ~\cite{j2}. Because of the invariance of the theory under the gauge transformations, this current is called as the conserved Noether current as $\partial_{\mu}J^{\mu}=0$. For a current localized in space, the Noether current is written as
\begin{equation}
\label{p11}
q_{sw}=\int{J^0 dV}=\int{\nabla . \left[\frac{1}{4 \pi} \frac{\hbar c}{\mathcal{V}_{sw} l^3}\right] dV} \,.
\end{equation}
The localized current in space then becomes
\begin{equation}
\label{p12}    
J^{\mu}=\frac{1}{4 \pi} \frac{3\hbar c}{\mathcal{V}_{sw} l^4}
\end{equation}
and leads to
\begin{equation}
\label{p1}
D_{\mu} G^{\mu \nu} = \frac{\alpha_{sw}}{g_{sw}} \frac{3\hbar c}{\mathcal{V}_{sw} l^4} \,.
\end{equation}
Here, we postulate that the $\mathcal{V}_{\nu}$ is the 4-potential for the strong and weak nuclear forces obeying
\begin{equation}
\label{q1}
G_{\mu \nu} = \partial_{\mu}{\mathcal{V}_{\nu}} - \partial_{\nu}{\mathcal{V}_{\mu}} + g \mathcal{V}_{\mu} \times \mathcal{V}_{\nu} \,.
\end{equation}
These are the non-abelian version of the emergent covariant Maxwell equations with the covariant derivative $D_{\mu}$.

\section{\label{sec:level7}Conclusions}

In this study, we attempt to describe all the fundamental interactions of nature as originated from an entropic behavior of the source and test items, under the influence of Freund's proposal ~\cite{c} based on the Verlinde's entropic gravity approach ~\cite{a}. For this purpose, we first start to investigate the entropic origin of the electromagnetic interaction and compare the results with the Wang's results ~\cite{j} in which he obtains a weird inconsistency in the physical significance of the thermodynamical interpretation by occurrence of negative temperature. We repair this inconsistency by taking the Planck energy $\hbar/t$ instead of the Wang's $Qc^2$ value that is in fact not a quantity defining an energy, for the energy value on the holographic screen causing the temperature $T$ on the screen. This energy and temperature on the screen due to the source charge are understood to induce an entropic electrostatic force ~\eqref{k} on a test charge approaching the screen from the emerged part of the spacetime due to the source charge. Furthermore, by generalizing the situation for a moving charge we obtain the emergent magnetic force ~\eqref{o} on a test current due to the source current inducing an energy and temperature on the screen enclosing the spacetime emerged by itself.

After obtaining the emergent forms of the Coulomb's and Biot-Savart laws, we generalize their entropic origin on electrostatic and magnetic fields by Gauss's and Ampere's law, respectively. As a direct result, we get the emergent nature of the classical Maxwell equations. Since the covariant relativistic form of the Maxwell equations lead to the abelian gauge field theories, we obtain emergent form of the covariant Maxwell equations which will provide a connection for the emergent nature of the non-abelian gauge field theories. From this connection, we postulate the emergent nature of the strong and weak forces with the color and isospin charge-based screen energy and temperature on a test charge, which leads to the entropic strong and weak interactions' field equations ~\eqref{o1}.

Consequently, the idea that the emergence of gravity and spacetime is entropic gives us some clues on all the fundamental forces of nature can be originated due to the entropy changes by certain type of entities, in certain type of limiting conditions after the big bang. This possibility makes us think that the universe, as a whole, can be considered as a gas of particles both in galactic scale and in quarks scale, since the observer's reference frame is possible to be taken to so far away from the galactic sample, so that the observer can infer as if they are Planck sized entities, relatively. Accordingly, considering the equivalence of the size scales of the entities of the universe through the observers relative reference position, it sounds quite sensible the entropic origination of the fundamental forces of nature which can be believed to be the descendant of a grand unified force.

\section{Conflict of Interest}

The authors declare that there is no conflict of interest regarding the publication of this paper.

\end{document}